\documentclass[11pt,letterpaper]{article} 

\usepackage[
  letterpaper,
  left=0.6in,right=0.6in,top=0.6in,bottom=0.6in,
  headheight=85pt,headsep=-45pt
]{geometry}
\usepackage{graphicx}
\graphicspath{{./}{figures/}}
\usepackage{float}
\usepackage{amsmath}
\usepackage{tikz}

\usepackage{caption}
\usepackage{booktabs}
\usepackage{array}
\usepackage{tabularx}
\usepackage{enumitem}
\usepackage{multirow}
\setlist{nosep}
\usepackage[table]{xcolor}
\definecolor{tableheader}{HTML}{D9D9D9}
\newcolumntype{P}[1]{>{\sffamily\centering\arraybackslash}p{#1}}
\newcolumntype{Y}{>{\sffamily\centering\arraybackslash}X}
\newcolumntype{A}[1]{>{\raggedright\arraybackslash}p{#1}}
\newcolumntype{M}[1]{>{\raggedright\arraybackslash}m{#1}}

\usepackage{listings}
\usepackage{xcolor}

\lstdefinelanguage{SysML}{
  morekeywords={part,def,attribute},
  sensitive=true,
  morecomment=[l]{//},
  morestring=[b]"
}
\lstdefinelanguage{diff}{
  morecomment=[f][\color{green!50!black}]{+},
  morecomment=[f][\color{red}]{-},
  morecomment=[f][\color{blue}]{@@},
}

\usepackage[T1]{fontenc}
\usepackage[utf8]{inputenc}
\usepackage{PTSerif}
\usepackage[scaled]{helvet}

\newcommand{\headingfont}{\sffamily}

\usepackage{titlesec}
\titleformat{\section}
  {\headingfont\bfseries\raggedright\fontsize{18pt}{18pt}\selectfont}{}{0.75em}{}
\titleformat{\subsection}
  {\headingfont\bfseries\raggedright\fontsize{15pt}{16pt}\selectfont}{}{0.75em}{}
\titleformat{\subsubsection}
  {\headingfont\bfseries\raggedright\fontsize{12pt}{14pt}\selectfont}{}{0.75em}{}

\titlespacing*{\section}{0pt}{7pt}{6pt}
\titlespacing*{\subsection}{0pt}{7pt}{6pt}
\titlespacing*{\subsubsection}{0pt}{7pt}{6pt}

\newcommand{\miniheading}[1]{%
  \par\noindent{\headingfont\bfseries\fontsize{12pt}{14pt}\selectfont #1}\par\vspace{4pt}%
}

\usepackage{fancyhdr}
\fancyhfoffset[R]{18pt}
\fancypagestyle{firstpage}{
  \fancyhf{}
  \fancyhfoffset[R]{18pt}
  \fancyhead[R]{%
    \smash{\raisebox{0pt}[0pt][0pt]{
      \begingroup\setlength{\fboxsep}{20pt}
        \colorbox{white}{\includegraphics[height=0.6in]{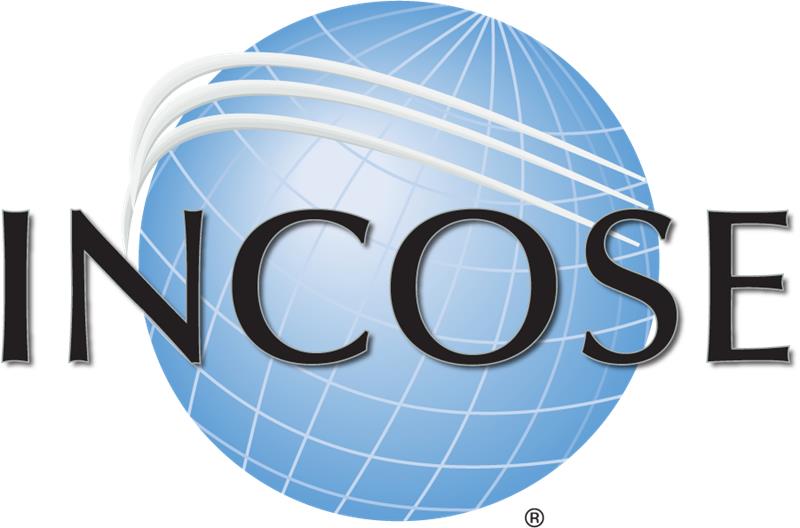}}%
      \endgroup
    }}
  }
  \fancyfoot[R]{\sffamily\bfseries\footnotesize \thepage}

}
\makeatletter
\@ifundefined{theauthor}{}{}
\makeatother
\usepackage{titling}
\pretitle{\headingfont\bfseries\fontsize{24pt}{26pt}\selectfont\raggedright}
\posttitle{\par\vspace{-.3in}}
\preauthor{}\postauthor{}
\author{\mbox{}}
\date{}
\newcommand{\authorcard}[5]{%
  {\headingfont\bfseries\fontsize{12pt}{14pt}\selectfont #1}\par
  {\headingfont\bfseries\fontsize{12pt}{14pt}\selectfont #2}\par
  {\headingfont\bfseries\fontsize{12pt}{14pt}\selectfont #3}\par
  {\headingfont\bfseries\fontsize{12pt}{14pt}\selectfont #4}\par
  {\headingfont\bfseries\fontsize{12pt}{14pt}\selectfont #5}\par
}



\makeatletter
\newcommand{\colfig}[2][]{%
  \IfFileExists{#2}{\includegraphics[width=\linewidth,#1]{#2}}{%
    \fbox{\parbox[b][1.5in][c]{\linewidth}{\centering \textit{Missing figure: }#2}}}%
}
\makeatother

\usepackage{csquotes}
\usepackage[style=apa,backend=biber]{biblatex}
\usepackage{changepage}

\usepackage{multicol}
\usepackage[hidelinks]{hyperref}

\title{Automated Semantic Fault Localization in SysML v2: A Human-in-the-Loop Framework Using Knowledge-Graph Augmented LLMs\\ }

\begin{document}
\maketitle
\thispagestyle{firstpage}


\noindent
\begin{tabular*}{\textwidth}{@{\extracolsep{\fill}} A{0.32\textwidth} A{0.32\textwidth} A{0.32\textwidth}}
  \authorcard{Haitham Al-Shami}{Aalto University}{Otakaari 1}{Espoo 02150, Finland}{haitham.al-shami@aalto.fi} &
  \authorcard{Rohail Malik}{Aalto University}{Otakaari 1}{Espoo 02150, Finland}{rohail.malik@aalto.fi} &
  \authorcard{Riku Ala-Laurinaho}{Aalto University}{Otakaari 1}{Espoo 02150, Finland}{riku.ala-laurinaho@aalto.fi} \\
  \multicolumn{3}{@{}c@{}}{\rule{0pt}{0.9\baselineskip}} \\[-0.2\baselineskip]
  \authorcard{Jari Vepsäläinen}{Aalto University}{Otakaari 1}{Espoo 02150, Finland}{jari.vepsalainen@aalto.fi} &
  \authorcard{Raine Viitala}{Aalto University}{Otakaari 1}{Espoo 02150, Finland}{raine.viitala@aalto.fi} 
\end{tabular*}
\addvspace{.75in}

\begin{multicols*}{2}
\raggedcolumns

{\headingfont\bfseries\fontsize{8pt}{12pt}\selectfont
Copyright~\textcopyright~ \the\year{} by the author(s). Permission granted to INCOSE to publish and use.}
\\
\phantomsection
\miniheading{Abstract}
Model-Based Systems Engineering (MBSE) methods that support formally defined modeling languages can provide feedback when models violate syntactic rules. However, semantic faults that preserve syntactic validity while violating domain-specific engineering constraints often remain undetected. SysML~v2 introduces a textual syntax for MBSE models in addition to its graphical notation, and this enables compiler-like validation of model structure and language conformance. However, because SysML~v2 is a modeling language, syntactic validity alone does not guarantee that the modeled system is semantically valid; the represented engineering relationships, such as mechanical, electrical, fluid, or signal connections, must also be physically meaningful. These domain-aware semantic errors can propagate through the design process and surface late as costly integration failures. This paper presents a human-in-the-loop framework for automated fault localization and repair suggestion in SysML~v2 models. This is done by combining a fine-tuned small language model with a domain knowledge graph that encodes physical compatibility rules between system elements. The knowledge graph guides the generation of synthetic training data by systematically introducing plausible domain violations, and augments the model at inference time to ground repair suggestions in valid engineering constraints. We demonstrate the framework using the vehicle systems domain, where the knowledge graph captures the relationships between the mechanical, electrical, fluid, and signal interfaces. The fine-tuned model is capable of capturing compiler-detectable syntax errors and the more subtle semantic violations that pass validation but violate domain physics. The model outputs unified diff patches that localize faults and present candidate repairs for engineer review; this preserves human judgment in the design process. Evaluation of 1,184 test samples shows that fine-tuning improves semantic fault repair from less than 3\% to more than 91\%, with patch-based output reducing token length by 50\%. The framework offers a practical path toward AI-assisted model verification that complements existing MBSE tools.

\phantomsection
\subsubsection{Keywords}
SysML v2, Model-Based Systems Engineering (MBSE), Large Language Models (LLMs), Semantic Fault Localization, Knowledge Graphs

\section{Introduction}
Model Based Systems Engineering (MBSE) is the practice of maintaining a formal, single, authoritative digital model of the system, from which all other views (documents, simulations) are derived. This model-centered approach supports early validation, traceability, and communication between different engineering disciplines. As systems grow in scale, explicitly modeling components, interfaces, requirements, and constraints helps engineers manage dependencies and maintain consistency more effectively than traditional document-based approaches.

SysML~v2 allows defining system structure, behavior, requirements, and constraints through formally specified textual notation alongside a graphical notation. This standardized textual syntax makes system models directly accessible to methods such as parsing, version comparison, automated analysis, and language model based repair. To this end, Large Language Models (LLMs) have been applied to SysML~v2 in three ways: as interactive assistants that integrate models from different organizations (\cite{li2025llm}), as reviewers that analyze models for potential issues (\cite{bouamra2025systemp}), and as autonomous agents that generate model fragments from natural language requirements (\cite{jin2025system, rafique2025enhancing}).

Despite these advances, SysML~v2 remains a modeling language in which the engineer’s intent plays a central role in shaping the model. Such intent is often implicit and difficult to capture within current LLM frameworks, which can limit their practical effectiveness. In addition, existing work highlights that domain-specific fine-tuning is constrained by the scarcity of high-quality SysML~v2 training data (\cite{jin2025system}). Consequently, general-purpose LLMs have limited exposure to valid SysML~v2 syntax and frequently produce syntactic errors or invalid constructs. Even when syntactically correct, models can create semantic hallucinations by generating elements that satisfy language rules and tool-level checks but violate engineering constraints. For example, a model may connect incompatible physical interfaces or assign units that are inconsistent with the represented quantities. Furthermore, context window limitations restrict an LLM’s ability to reason over large systems, where relevant constraints may span thousands of lines.

In contrast to software engineering, where executable tests provide feedback on whether code satisfies its intended behavior, SysML~v2 models are not primarily executable artifacts. Their correctness depends on accurately representing design intent and adhering to domain-specific constraints. A generic test-based approach is therefore insufficient, as constructing tests that fully encode the intended system design would require much of the domain knowledge the model is expected to capture. This limitation highlights the importance of incorporating domain semantics and engineering judgment into automated MBSE methods.

\begin{figure*}[t]
  \centering
  \colfig{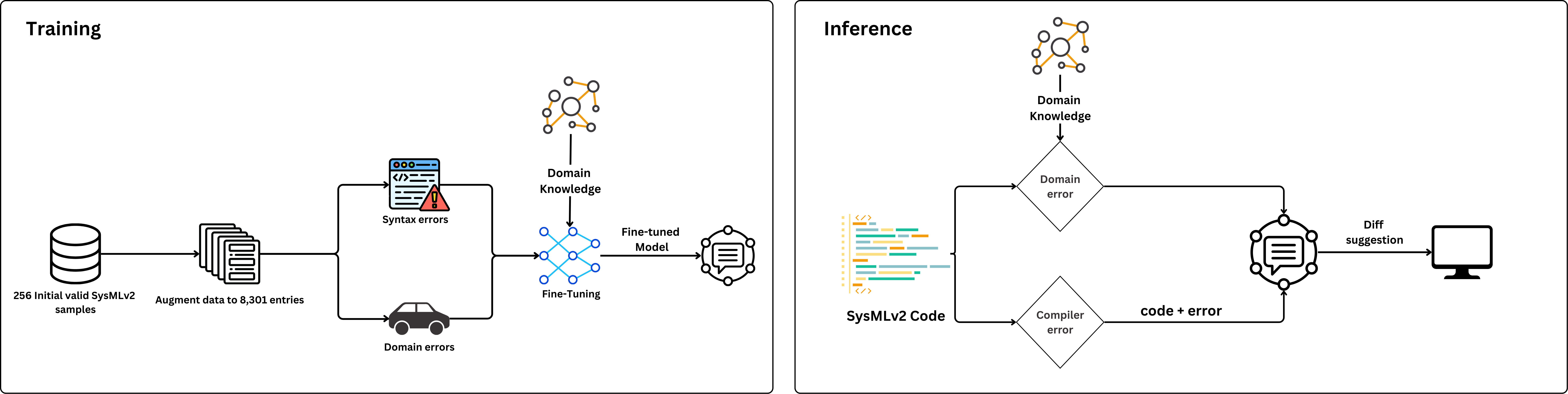}
  \caption{Overview of the proposed framework. A base corpus of 256 valid SysML~v2 samples is augmented through heuristic mutation to produce 8,301 entries comprising syntax errors, domain errors, and correct examples. A small language model is fine-tuned on this data with domain knowledge graph augmentation. At inference, input code undergoes compiler validation and domain constraint checking, with the model outputting diff patch suggestions for engineer review.}
  \label{fig:pipe_line}
\end{figure*}

In this paper, we thus apply language models for fault correction instead of model generation. To address the scarcity of SysML~v2 data, we construct a dataset containing two classes of faulty models: syntactic errors, which violate SysML v2 grammar rules and fail compilation, and domain errors, which are syntactically valid but semantically incorrect and undetectable by compilers. Using this dataset together with a domain-rules knowledge graph, we investigate the effectiveness of fine-tuned small language models (SLMs) for fault identification and correction across both types of errors. The models are trained to generate unified diff patches instead of full code rewrites, allowing for efficient human-in-the-loop repair workflows. Figure \ref{fig:pipe_line} shows the method employed in this paper which includes both training and inference. The code for data synthesis and fine-tuning methodology utilized in this paper is publicly available\footnote{https://github.com/rohailamalik/SysMLv2-repair-with-KG-SLMs}. 

The primary contributions of this paper are:
\begin{itemize}
    \item A dataset of correct, syntactically, and semantically faulty SysML~v2  models, and its synthesis methodology.
    \item A domain-aware fault localization pipeline guided by a knowledge graph of physical interface and quantity constraints.
    \item Evaluation of two fine-tuned SLMs for identifying and repairing syntactic and semantic SysML~v2 faults through unified diff outputs.
\end{itemize} 

\section{Related Work} 
This section reviews literature that integrates LLMs with SysML~v2 for model generation, evaluation metrics, data preparation, and model validation. \textcite{cibrian2025agent} examined the use of general-purpose LLMs for generating SysML~v2 models and report that direct prompting often produces outputs that appear structurally plausible but fail syntactic validation. Common failure modes include hallucinated SysML~v1 syntax and nonexistent SysML~v2 keywords. To address these issues, the authors propose SysMLAgent, an autonomous agent-based framework that decomposes model generation into iterative planning, code generation, validation, and repair steps. The framework uses a Belief-Desire-Intention architecture to maintain the current model state, pursue the goal of producing a valid SysML~v2 model, and update its actions based on validation feedback. They integrate Retrieval-Augmented Generation (RAG), ANTLR validation, and a generation loop to limit issues of hallucinations and improve the generated model quality. Their agentic architecture achieved a score of 2.5 on their evaluation benchmark, whereas baseline LLMs achieved 1.0. 

\textcite{cibrian2025agent} primarily addressed syntactic validity rather than domain-level semantic correctness. Parser-based validation can identify grammar violations, but cannot determine whether a generated model satisfies physical compatibility constraints or reflects valid engineering intent. In addition, the framework relies on retrieval from a limited example database, which may reduce robustness when user requests fall outside the covered modeling patterns. These limitations motivate approaches that combine language models with explicit domain knowledge.

\textcite{DeHart2024} presented an LLM-driven approach for interacting with SysML~v2 models through the standard SysML~v2 API using natural language input. In this workflow, the LLM is given access to an existing ground-truth model and is tasked with performing both simple and complex model-editing operations. When the generated code is not compiler-ready, the system uses the resulting runtime exception or syntax error as feedback in a regeneration loop, allowing the LLM to iteratively revise its output through in-context learning. The study identifies several limitations of LLM-driven system design approaches. First, natural-language requests can be ambiguous, leading the LLM to misinterpret the engineer’s intent. Second, direct modification of SysML~v2 code may fail to preserve the syntactic and semantic precision required by the language. Third, generated code and model edits require robust error handling, rollback, and validation to avoid corrupting the model or producing technically unreliable results. Fourth, output quality depends strongly on the quality, currency, and completeness of the contextual model information provided to the LLM. The paper also notes security and privacy concerns, workflow-integration costs, and the risk of automation bias or overreliance on generated outputs. These findings highlight the need for explicit, well-structured domain knowledge and independent validation when applying LLMs to SysML~v2 modeling tasks.

Incorporating Knowledge Graphs (KG) into LLM generation of SysML~v2 has been studied. \textcite{Qualis25} built a tri-layered (KG) consisting of a SysML~v2 layer, a domain-specific layer and a system specific layer for the model being built. Automatic parameter injection as prompt templates to the LLM; the code is validated against the official SysML~v2 API, and in the case of errors, a one-pass self-repair is initiated for the LLM. This method proved effective in mitigating hallucinations. However, issues of limited dataset, semantic fidelity (how well the model captures domain concepts) remain a challenge. LLMs struggled with abstract or nuanced concepts, and semantic scores were consistently lower than structural ones. The author acknowledges expert bias in the evaluation phase; the ground truth dataset itself was created by the same single expert. This means the expected answers the LLMs are being tested against are inherently biased toward one person's modeling style, rather than a  standard benchmark.

\textcite{jin2025system} introduced SysMBench, a corpus of 151 design scenarios, showing that state-of-the-art models achieve low pass rates on complex system descriptions due to a scarcity of training data. To mitigate this, \cite{rafique2025enhancing} compared general-purpose reasoners (GPT-4) against fine-tuned code models (CodeT5), finding that while fine-tuning improves syntactic compliance, it often fails to correct deeper logical inconsistencies. Alternative architectural approaches have attempted to constrain this uncertainty; \textcite{bouamra2025systemp} proposed SysTemp, a multi-agent system that enforces structural validity through rigid templates, though this limits the expressivity needed for novel designs. Similarly, \textcite{li2025llm} explored LLM-assisted semantic alignment for collaborative modeling, utilizing iterative human verification to map disjoint system models, further highlighting that human oversight remains a critical dependency in current AI-MBSE workflows.

In response to these challenges, this paper presents a framework that shifts focus from autonomous model generation to automated semantic fault localization. Addressing the data scarcity highlighted by \textcite{jin2025system} and \textcite{Qualis25}, we introduce a methodology for synthesizing a large-scale dataset containing syntactic and semantic faults. The latter are syntactically valid system models, but violate domain physics, and are generated by augmenting scarce samples with knowledge-graph-derived corruptions. Our architecture couples a fine-tuned LLM with a rule-based RAG system, allowing the model to detect these subtle errors and ground its repair suggestions in established engineering constraints. In addition, the models are trained to output diff patches instead of entire repaired code. This approach positions the LLM as a domain-aware assistant rather than an autonomous agent, preserving engineer oversight while grounding repair suggestions in explicit domain constraints.

\section{Methodology}
This section presents the workflow in the order of its construction and use. The~\nameref{sec:domainkg} subsection introduces a knowledge graph (KG) formalizing physical compatibility constraints which are used for both dataset generation and prompt augmentation. The \nameref{subsec:Dataset_synthesis} subsection then describes how a small corpus of valid SysML~v2 models is expanded into a labeled dataset through \nameref{sec:syntax-fault}, \nameref{sec:kg-semantic-fault}, and inclusion of correct examples for error classification (\nameref{subsec:correct_examples}).

The \nameref{subsec:Patch-based_outputs} subsection motivates the use of unified diff patches as the model output format. Finally, \nameref{subsec:Supervised_Fine-tuning} subsection outlines the fine-tuning procedure, including data-leak-proof split (\nameref{subsec:dataset_split}), the evaluation methods (\nameref{subsec:evaluation}), and the selection of base models (\nameref{subsec:model-selection}).

\subsection{Domain Knowledge-Graph} 
\label{sec:domainkg}
The knowledge graph (KG) encodes the physical and engineering constraints used to evaluate SysML~v2 models in the target domain and accomplishes two roles in the proposed framework. 
\begin{enumerate}
    \item Supporting the mutation engine by providing domain rules which are violated to generate semantically faulty training examples.
    \item Augmenting the prompts with rules relevant to the code being analyzed, providing domain context for semantic fault detection and repair.
\end{enumerate} 

\begin{figure*}[t]
  \centering
  \colfig{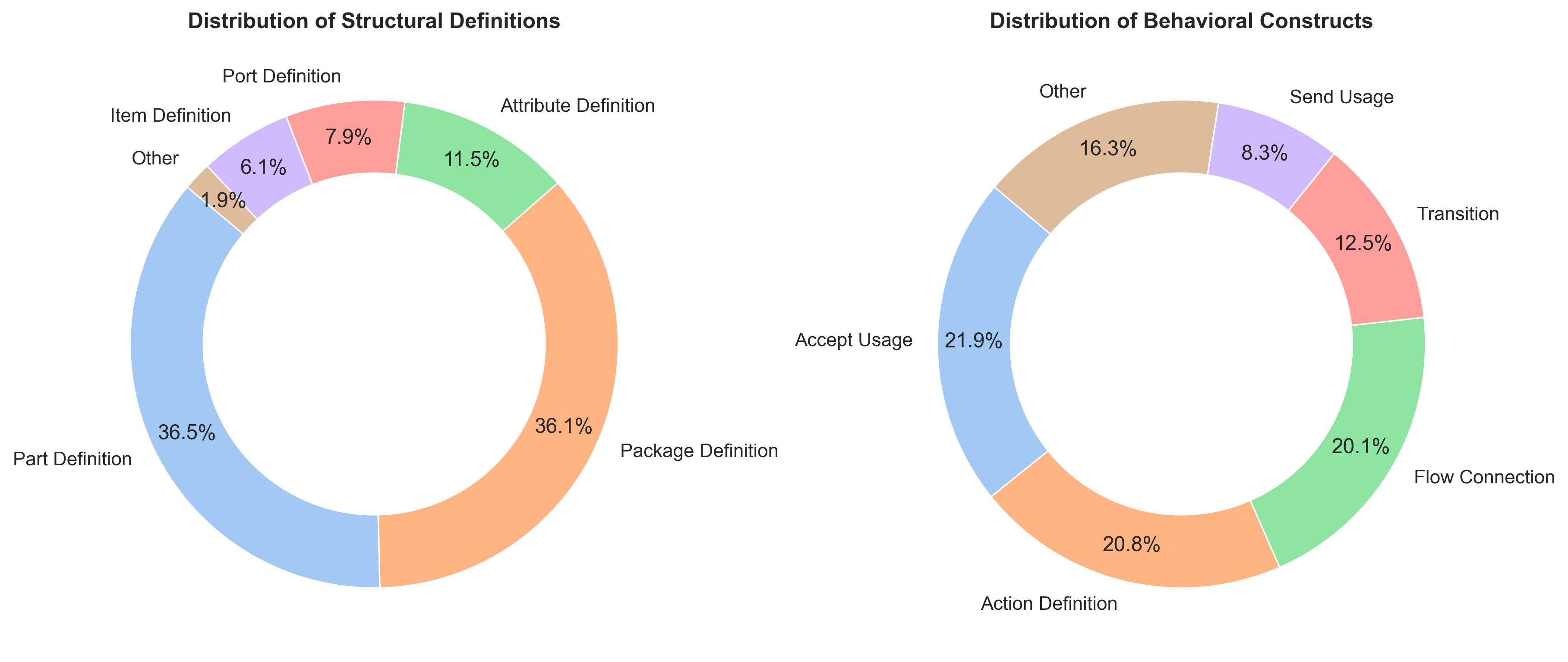}
  \caption{Breakdown of the structural and behavioral elements of the SysMLv2 code in the dataset}
  \label{fig:dataset_stats}
\end{figure*}

In this work, we evaluate the approach in the vehicle systems domain. The domain knowledge is represented using two lightweight graphs:
\begin{itemize}
    \item \textbf{Vehicle Domain KG:} This graph maps system interfaces, such as \texttt{DrivePwrPort} and \texttt{FuelPort}, to high-level physical domains, such as \texttt{MECHANICAL\_TORQUE} and \texttt{FLUID\_FUEL}. It enforces an adjacency constraint in which connections are permitted only between ports belonging to compatible physical domains. The vehicle-domain concepts are derived from the base dataset of 256 valid SysML~v2 models.
    
    \item \textbf{Physical Quantity KG:} This graph maps International System of Quantities (ISQ) types to valid units. It is used to identify dimensional inconsistencies, such as assigning voltage units, \texttt{[V]}, to a mass quantity.
\end{itemize}

\subsection{Dataset Synthesis}
\label{subsec:Dataset_synthesis}
The baseline dataset was constructed using the official SysML~v2 Pilot Implementation repository (\cite{SysMLv2Pilot2024}) to address the scarcity of high-quality SysML v2 corpora. This repository provides examples of syntactically and semantically valid models.

The dataset contains 256 unique SysML v2 files, organized into domains such as vehicle systems, geometry, mass roll-ups, and state-space representations. Static analysis shows a distribution bias toward structural modeling elements. As shown in Figure \ref{fig:dataset_stats}, part usage and attribute usage account for nearly 39\% of all identified constructs, reflecting the language's primary use in defining system architectures. Behavioral constructs (e.g., state def) and requirements appear less frequently, which poses a challenge for training LLMs to learn dynamic system behaviors.

\subsubsection{Synthetic Syntax Fault Generation}
\label{sec:syntax-fault}
To train the model to recognize and repair syntactic violations, we developed a mutation engine that systematically injects errors into valid SysML~v2 models. A set of heuristic mutations is applied to each of the 256 SysML~v2 files in the dataset. The process of producing syntactic faults is defined as follows:
\begin{itemize}
    \item The valid SysML v2 source file is loaded.
    \item A specific mutation operator is applied to create a potentially corrupted candidate.
    \item The candidate is executed against a SysML~v2 Jupyter kernel (via Jupyter client).
    \item If the kernel returns a compilation error, the pair (Broken Code, Error Message) is saved. If the mutation accidentally results in valid code (e.g., renaming an unused variable), the sample is discarded.
\end{itemize}
This process was applied to the source dataset, which yielded 5,497 verified syntactic fault samples. The 12 distinct heuristics are listed and shown in Table \ref{tab:mutation_taxonomy}.

\begin{table*}[t]
\renewcommand{\arraystretch}{1.3}
\begin{tabularx}{\textwidth}{@{}l l X@{}}

\rowcolor{tableheader}
\multicolumn{1}{c}{\headingfont\bfseries Category} &
\multicolumn{1}{c}{\headingfont\bfseries Operator} &
\multicolumn{1}{c}{\headingfont\bfseries Description} \\
\addlinespace[8pt]

\multicolumn{3}{c}{\headingfont\bfseries Synthetic Mutation Operators} \\
\addlinespace[6pt]

\multirow{3}{*}{Syntax \& Grammar}
& \textit{Break Imports} & Corrupts standard library imports (e.g., \texttt{::*;} $\to$ \texttt{::;}) \\
& \textit{Missing Semicolon} & Removes terminal semicolons from random lines \\
& \textit{Unbalanced Brackets} & Deletes random closing brackets to break scoping \\
\addlinespace[8pt]

\multirow{3}{*}{Visibility \& Scope}
& \textit{Flip Visibility} & Swaps \texttt{public}/\texttt{private} keywords to force access errors \\
& \textit{Drop Import} & Removes required import lines to trigger unresolved symbols \\
& \textit{Duplicate Feature} & Duplicates attribute definitions to force name collisions \\
\addlinespace[8pt]

\multirow{3}{*}{Type \& Units}
& \textit{Unit Mismatch} & Injects incompatible units (e.g., $[N \cdot m] \to [kg]$) \\
& \textit{Drop Units} & Removes required unit definitions from quantities \\
& \textit{Flip Port} & Inverts port conjugation (e.g., $\sim$DriveIF $\to$ DriveIF) \\
\addlinespace[8pt]

\multirow{3}{*}{Logic \& Consistency}
& \textit{Multiplicity Inversion} & Sets lower bound greater than upper bound (e.g., $[1..0]$) \\
& \textit{Typo Endpoint} & Alters connection targets to mimic spelling errors \\
& \textit{Keyword Swap} & Swaps semantic pairs such as \texttt{subsets}/\texttt{redefines} \\

\addlinespace[12pt]
\hline
\addlinespace[12pt]

\multicolumn{3}{c}{\headingfont\bfseries Semantic Mutation Operators} \\
\addlinespace[6pt]

\multirow{2}{*}{Interface Physics}
& \textit{Port Mismatch} & Connects incompatible ports (e.g., Torque $\leftrightarrow$ Electrical) \\
& \textit{New Bad Connection} & Creates a new wire between disparate physical domains \\
\addlinespace[8pt]

\multirow{3}{*}{Quantity Physics}
& \textit{Unit Swap} & Swaps valid units for incompatible ones (e.g., $kg \to m$) \\
& \textit{Kind Mismatch} & Changes attribute kind (e.g., Mass $\to$ Length) \\
& \textit{Compound Break} & Corrupts derived units (e.g., $[N \cdot m] \to [N / m]$) \\

\end{tabularx}
\caption{Taxonomy of Synthetic and Semantic Mutation Operators}
\label{tab:mutation_taxonomy}
\end{table*}

\subsubsection{KG Guided Semantic Fault Generation} 
\label{sec:kg-semantic-fault}
Standard compilers detect syntax errors, but semantic faults, such as connecting a fluid interface to an electrical port, require domain-specific logic to identify. As such, we utilize the rules encoded in the domain knowledge graph to generate training data for these latent errors. The engine systematically injects violations of physical and engineering constraints into otherwise valid models.

The engine parses SysML~v2 models to construct an in-memory property graph of all part definitions, instances, and ports. It then applies five semantic mutation heuristics (see Table \ref{tab:mutation_taxonomy}) to alter the graph topology or metadata violating the KG rules.

A requirement for our synthetic data is that semantic mutations remain syntactically valid. To ensure the mutations are useful for training, every generated model must pass the SysML~v2 compiler. The pipeline discards any mutation that triggers a syntax error, retaining only those that compile successfully but represent logical or physical impossibilities.

The mutation process yielded 1,402 verified semantic fault samples. The distribution is heavily weighted toward interface mismatch errors, which account for 93.1\% of the total dataset: 1,077 samples (76.8\%) resulted from mismatches between existing ports, while 228 (16.3\%) involved the generation of new invalid connections. The remaining faults comprised quantity and unit-level inconsistencies, including Quantity Kind mismatches (6.1\%, $n=86$), Unit Incompatibilities (0.6\%, $n=8$), and Compound Unit Errors (0.2\%, $n=3$). 

\begin{figure*}[t]
  \centering
  \colfig{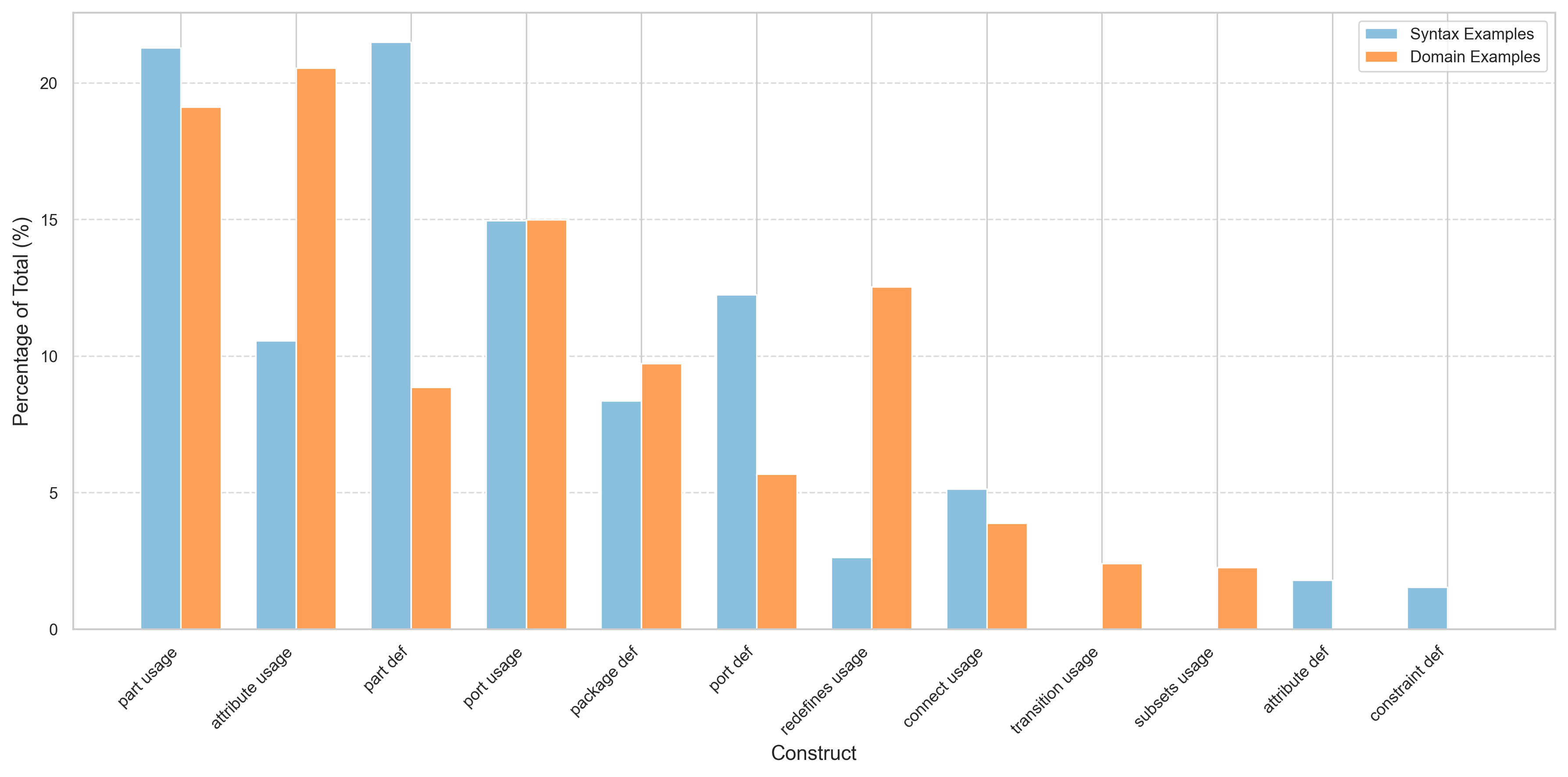}
  \caption{Representation of different constructs in syntax and domain/semantic error examples}
  \label{fig:bar_comparison}
\end{figure*}

This distribution is a direct consequence of the mutation engine's heuristic yield and the structural density of the source models. Because the models are primarily defined by their port-based topology, the search space for connectivity violations is significantly larger than that for physical quantity assignments. Figure \ref{fig:bar_comparison} compares the data breakdown for domain and syntax error sets. Generally, both sets contain comparable constructs with some skews due to the construction depending on runtime exception. This discrepancy impacts the ability of the model and its consistency given error types. 

\subsubsection{Correct Examples} 
\label{subsec:correct_examples}
In practice, syntax errors are always indicative of a mistake, as they are detected by the compiler and violate language constructs. In contrast, semantic errors are not explicitly identifiable, which extends the role of the language model from code repair to error classification. To enable the model to learn the distinction between semantically valid and invalid code, we include examples of code that is both syntactically and semantically correct. These examples correspond to the base instances used to generate the rest of the dataset and are duplicated to match the number of semantically invalid samples.
This balanced dataset allows the model to classify whether given code is correct and repair it when it is not.

The final dataset thus comprises 5,497 syntax fault examples, 1,402 semantic fault samples, and 1,402 correct code samples, producing an overall dataset of 8,301 entries. Each sample contains ground truth Good Code and mutated Code. Compiler Errors are appended to Syntax error entries, while the rest (semantic-error and correct examples) contain relevant domain rules from the knowledge graph.

\subsection{Patch-based outputs} 
\label{subsec:Patch-based_outputs}
To support a human-in-the-loop repair, the model is fine-tuned to generate code patches rather than full repaired code. This approach also reduces output length, which lowers computational cost during inference. The resulting patches can be reviewed or modified by a human user, and applied manually or automatically.

The patches are in the form of unified diff patches, which specify code changes through added and removed lines. In unified diffs, changes are localized using unchanged context lines before and after the modification, as shown in Listing \ref{lst:sysml-diff}. While line numbers do appear in the header, they are an initial reference for patch application. The actual anchoring relies on matching context lines, making the approach robust even when line numbers are inaccurate.

This context-based anchoring is advantageous because precise line-number tracking is difficult for language models given their token-prediction architecture. Injecting explicit line numbers into the code is possible, but it introduces noise and can obscure code semantics. However, generating patches instead of full code also requires more structured reasoning, as the model must identify relevant context, follow the diff format, and correctly express edits.

\begin{lstlisting}[language=diff, caption={Example SysML patch in unified diff format}, label={lst:sysml-diff}]
@@ -1,5 +1,5 @@
 package Vehicle_Remix_0190_6c85 {
-    port def ControlPort
+    port def ControlPort;
     port def LugNutPort;
     port def LugNutCompositePort;
     part def LugNutPort_Def { port p : LugNutPort; }
\end{lstlisting}

\subsection{Supervised Fine-tuning}
\label{subsec:Supervised_Fine-tuning} 
The code repair is formulated as a supervised learning task given input code \( c \) and an associated signal \( s \) (error description for syntax errors, and domain rules for semantic errors or correct examples).
The output is \( z \in \{\texttt{correct}, \texttt{incorrect}\} \) i.e., a code status indicator, and the repair output \( r \), which may be empty, a full repaired code sequence, or a unified diff patch.
The learning objective is thus to model the conditional distribution
\[
p_\theta(y \mid x) = p_\theta(z, r \mid c, s).
\]

As the model is a causal language model, the conditional distribution is factorized auto-regressively as:
\[
p_\theta(z, r \mid c, s)
=
\prod_{t=1}^{T}
p_\theta(y_t \mid c, s, y_{<t}),
\]
where \( y = (y_1, \ldots, y_T) \) is the output token sequence, with the status token \( z \) generated before the repair content \( r \). 

Fine-tuning is performed using Low-Rank Adaptation (LoRA), which updates only a small subset of model parameters while keeping the rest frozen. This enables task adaptation while preserving the pretrained model weights and reducing the number of trainable parameters. During training, the model undergoes teacher forcing, where it learns next-token prediction given ground-truth context at each token, and optimization is driven by token-level loss. Early stopping is applied based on evaluation loss computed on a held-out validation set, also using teacher forcing. 
The prompts and completions used for fine-tuning and evaluation are provided in \nameref{app:prompts_completions}.

\subsubsection{Dataset Split}
\label{subsec:dataset_split}

For the model to effectively generalize instead of memorizing the training examples, the training, evaluation, and test datasets are kept disjoint. To prevent data leakage, source examples from which mutated samples are generated are unique across splits. The splitting procedure also ensures that syntactic and semantic mutation types are uniformly distributed across all datasets.

Instead of truncating longer examples, which can degrade learning in code repair tasks due to partial code inputs, we drop examples exceeding 2048 tokens (prompt plus response).
This filtering removes 521 examples, and the remaining data is split using a 70:15:15 ratio.
This results in 5,451 training, 1,184 evaluation, and 1,145 test examples.

\subsubsection{Evaluation}
\label{subsec:evaluation}

After training, the model is evaluated on a held-out test set using auto-regressive generation to mirror inference-time conditions. Outputs are produced using greedy decoding, where the highest-probability token is selected at each step. This choice is motivated by the largely deterministic nature of code repair, particularly for syntax error cases.

Each test instance is passed to the model once, and the generated patches are applied to the faulty code. The resulting repaired code is compared against the corresponding ground-truth solution. 
Compiler-based evaluation is not used for two reasons.

Correctness is assessed through strict string-level comparison with the ground truth. Generated outputs are normalized by removing comments and applying consistent formatting before comparison. A prediction is marked as correct only if it exactly matches the reference solution.  
This strict criterion is necessary because even minor discrepancies, such as a missing brace, can leave syntactic errors unresolved despite high semantic similarity.

\subsubsection{Model Selection}
\label{subsec:model-selection}

To assess the robustness of the proposed approach across model sizes and architectures, we evaluate SLMs of varying scale. 
All selected models are coder models, based on the assumption that models pretrained on programming languages better capture programming structure and repair heuristics than general-purpose chat models.
Specifically, we evaluate Qwen-2.5 Coder 1.5B Instruct (\cite{hui2024_qwen2}) and DeepSeek Coder 6.7B Instruct (\cite{guo2024_deepseekcoder}).

\begingroup
\begin{table*}[t]
\renewcommand{\arraystretch}{1.3}
\begin{tabularx}{\textwidth}{@{}l l *{4}{>{\centering\arraybackslash}X}@{}}
\rowcolor{tableheader}
\multicolumn{1}{c}{\headingfont\bfseries Model} &
\multicolumn{1}{c}{\headingfont\bfseries Type} &
\multicolumn{1}{c}{\headingfont\bfseries Baseline} &
\multicolumn{1}{c}{\headingfont\bfseries With rules} &
\multicolumn{1}{c}{\headingfont\bfseries FT-Code} &
\multicolumn{1}{c}{\headingfont\bfseries FT-Patch} \\
\addlinespace[8pt]

\multirow{5}{*}{Qwen2.5 Coder 1.5B Instruct}
 & Synthetic & 17.6\% & 17.6\% & 90.7\% & 62.9\% \\
 & Semantic  &  0.62\% &  1.85\% & 95.7\% & 91.9\% \\
 & None      & 54.4\% & 62.6\% & 98.1\% & 98.1\% \\
 & Overall   & 21.8\% & 23.5\% & 92.8\% & 73.4\% \\
 & Output Tokens & 316 & 262 & 173 & 84 \\
\addlinespace[10pt]

\multirow{5}{*}{DeepSeek Coder 6.7B Instruct}
 & Synthetic     & 36.1\% & 36.1\% & 89.3\% & 57.9\% \\
 & Semantic      & 2.47\% & 0.29\% & 91.9\% & 91.4\% \\
 & None          & 28.6\% & 29.1\% & 94.7\% & 97.6\% \\
 & Overall       & 30.0\% & 29.8\% & 90.7\% & 70.2\% \\
 & Output Tokens & 345    & 388 & 271    & 111 \\

\end{tabularx}
\caption{pass@1 scores across model variants. FT-Code and FT-Patch stand for models fine-tuned to output correct code and difference patch, respectively. Baseline involves passing semantic error examples and correct examples without appending relevant domain rules, while with rules involves their explicit addition to the prompt through knowledge graph. Rules are appended in both fine-tuned variants. The output tokens for each configuration are averaged over the test dataset}
\label{tab:llm-performance}
\end{table*}
\endgroup

For benchmarking, all models used are evaluated on the test set, both with and without domain rules provided for semantic error cases. 
This comparison isolates the impact of domain rules on semantic repair performance independent of fine-tuning. 
Finally, we also fine-tune the models to generate repaired code in its entirety to allow a direct comparison against the patch-outputting model and corresponding reasoning demands.
All models are fine-tuned on a single NVIDIA A100 GPU (80 GB VRAM), and set to train over 3 epochs. 

\section{Results}
This section presents the performance of the fine-tuned models on syntactic and semantic error correction tasks. Table \ref{tab:llm-performance} reports pass@1 scores for all evaluated models and output variants across both error types, together with the average number of output tokens on the test dataset.

Providing domain rules in the prompt yields only marginal performance gains for semantic error cases and correctly formed inputs, whereas supervised fine-tuning results in substantial improvements over both baseline settings. 
Models trained to generate patch-based outputs achieve slightly lower accuracy than those generating full repaired code.
This reflects the increased reasoning complexity of structured patch generation.

Across all configurations, performance improves with increasing model size. 
At the same time, patch-based models reduce output length by approximately 50\% compared to fine-tuned full-code output models, and by up to 75\% relative to baseline models.

\begin{figure*}[t]
  \centering
  \includegraphics[width=1\textwidth]{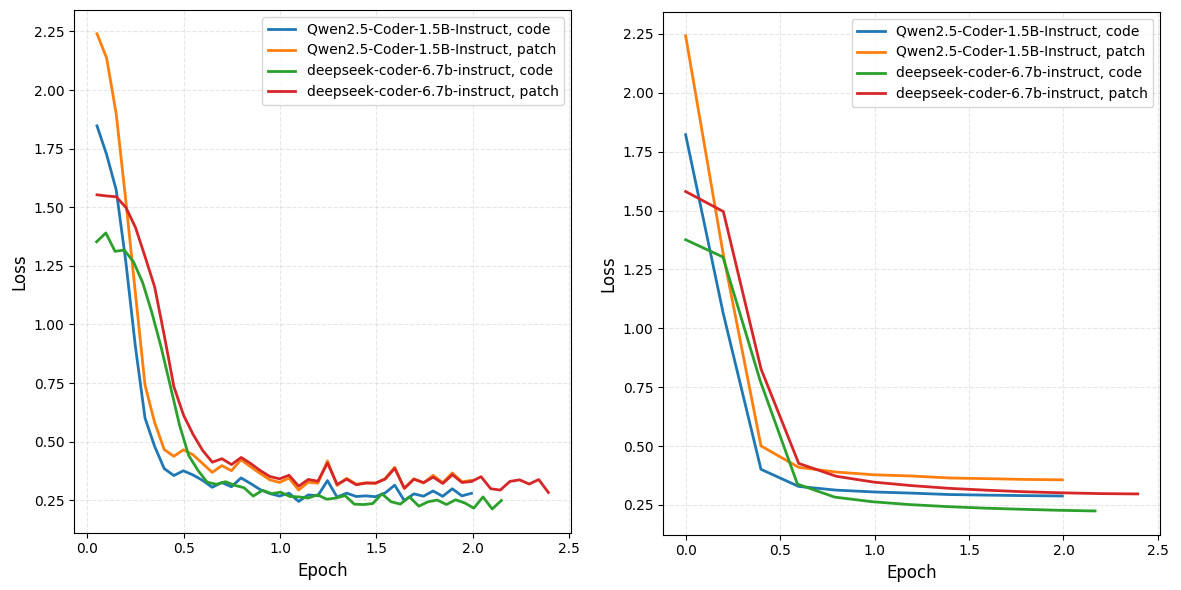}
  \caption{Losses over training (left) and evaluation (right) datasets during training. Training is set to last 3 epochs, but is early stopped due to evaluation loss convergence}
  \label{fig:train_eval_losses}
\end{figure*}

Listing \ref{lst:bad-code} shows an example input of semantically faulty code with correct syntax. Although this code is syntactically correct and would pass a standard SysML v2 compiler, it violates physical constraints.

\begin{lstlisting}[language=SysML, caption={Input Snippet with Semantic Error}, label={lst:bad-code}]
    part def MassedThing {
        attribute mass :> ISQ::mass; 
        attribute totalMass :> ISQ::force; // Semantic Error
    }
\end{lstlisting}

The model processes this input and generates a unified diff patch. As shown in Listing \ref{lst:sysml-diff-fix}, the patch correctly localizes the fault and replaces the invalid \texttt{force} type with the \texttt{mass} type required by the Ground Truth reference shown in Listing \ref{lst:good-code}.

\begin{lstlisting}[language=diff, caption={Model-Generated Patch}, label={lst:sysml-diff-fix}]
@@ -2,7 +2,7 @@
 	part def MassedThing {
 		attribute mass :> ISQ::mass; 
-		attribute totalMass :> ISQ::force;
+		attribute totalMass :> ISQ::mass;
 	}
\end{lstlisting}

\begin{lstlisting}[language=SysML, caption={Ground Truth Reference Snippet}, label={lst:good-code}]
    part def MassedThing {
        attribute mass :> ISQ::mass; 
        attribute totalMass :> ISQ::mass;
    }
\end{lstlisting}

Figure \ref{fig:train_eval_losses} shows the training and evaluation loss curves over epochs for each fine-tuned model configuration. 
For all models, early stopping is triggered once the evaluation loss converges.

\section{Discussion}
The core idea of this paper is to shift the application of LLMs in systems engineering from autonomous model generation to automated fault localization and repair, specifically targeting semantic errors that bypass standard compilers. A Small Language Model (SLM) was fine-tuned with rule-based domain knowledge for a human-in-the-loop workflow through the use of diff patches. 

The significant improvement in performance after fine-tuning suggests that while pre-trained coder models have a high structural literacy, they require specific adaptation for the semantic logic of SysML v2. The rapid convergence of evaluation loss, as seen in Figure \ref{fig:train_eval_losses}, indicates that the models successfully mapped the domain rules from the knowledge graph to the specific syntax of the system models.

There is an observed trade-off between output efficiency and reasoning complexity; although generating diff patches reduces computational costs and improves human-in-the-loop utility, it introduces a reasoning penalty that slightly lowers accuracy compared to full code rewrites. Furthermore, the high success rate in semantic repair demonstrates that the framework effectively generalizes from the structural patterns in the training data—which is primarily composed of port and connection logic to identify latent physical impossibilities that standard compilers are unable to detect.

The current pipeline has several limitations and constraints. The primary constraint lies in the relatively limited augmentation types with 12 syntactic and 5 semantic mutation heuristics. Because these synthetic faults are generated systematically, they may not fully capture the variety of organic errors produced by human engineers in complex, real-world industrial environments. Although the presented dataset is a step toward more accessible data for SysML~v2 code LLM usage, more real-world data is still required. 

In semantic fault repair, multiple valid engineering solutions may exist for a single violation. In this framework, the model's repair logic is governed by the constraints encoded in the Knowledge Graph. This deterministic approach is advantageous because it ensures that all suggestions are grounded in expert-verified domain knowledge and physical laws. However, this rigid adherence also introduces a trade-off in design flexibility of the LLM-suggested fixes. 

The poor baseline performance of pre-trained models on SysML~v2 confirms observations by \textcite{cibrian2025agent}, who reported that raw LLMs frequently hallucinate syntax from SysML~v1 or fabricate nonexistent constructs. Similarly, the baseline semantic repair rate of under 3\%  aligns with \textcite{Qualis25}, who  found that semantic scores remained consistently lower than structural ones even with knowledge graph augmentation. However, the fine-tuning results in \textcite{rafique2025enhancing} concluded that fine-tuning improves syntactic compliance but fails to correct deeper logical inconsistencies. This work demonstrates that when training data explicitly encodes domain violations through knowledge graph guided mutations, fine-tuned models can achieve over 90\% accuracy on semantic fault repair. The emphasis on human-in-the-loop workflows through diff patches aligns with \textcite{li2025llm}, who found that human oversight remains a critical dependency in current AI-MBSE systems. Finally, the synthetic data generation methodology improves the data scarcity problem highlighted by \textcite{jin2025system}, whose SysMBench study attributed low model performance to limited SysML~v2 training corpora.

On the theoretical side, the heuristic-based data augmentation methodology demonstrates a viable path for overcoming the scarcity of SysML~v2 training data. Although the current implementation uses a limited set of mutation operators applied to a small source corpus, which may affect generalizability, the approach establishes that synthetic fault injection guided by domain knowledge can produce effective training data where organic examples are unavailable. The resulting dataset of over 8,000 annotated samples itself represents a resource for future research in AI-assisted systems modeling.

On the practical side, the human-in-the-loop design addresses hallucination risks by positioning the language model as an assistant rather than an autonomous agent, ensuring that engineers retain authority over modeling decisions. This extends the utility of language models beyond syntax checking to detecting domain-level violations that may otherwise go unnoticed in complex system architectures. Furthermore, because the same knowledge graph governs both training data generation and inference-time augmentation, the framework enforces consistency across project files, promoting uniformity in design decisions. As a system matures, the knowledge graph can be extended to encode additional domain constraints, allowing engineering teams to accumulate institutional knowledge in a form that directly enhances model-assisted verification throughout the system lifecycle.

\section{Conclusion}
This paper presented a human-in-the-loop framework for automated semantic fault localization and repair in SysML~v2 models. The approach combines fine-tuned small language models with a domain knowledge graph that encodes physical compatibility constraints between system elements. To address the scarcity of SysML~v2 training data, a heuristic-based mutation engine was developed to synthesize over 8,000 faulty and correct code samples, covering both compiler-detectable syntax errors and latent semantic violations. The fine-tuned models were trained to output unified diff patches rather than full code rewrites, supporting efficient engineer review. Evaluation on a held-out test set showed that fine-tuning improved semantic fault repair accuracy from under 3\% to over 90\%, while patch-based outputs reduced token length by approximately 50\% compared to full code generation. These results demonstrate that knowledge-graph-guided data augmentation can overcome training data limitations, and that small language models, when properly adapted, can detect domain-level errors that standard compilers miss. The framework positions the language model as an assistant rather than an autonomous agent, preserving engineering judgment while extending verification capabilities beyond syntax to domain physics.

\section{Acknowledgments}
This research was partly funded by Business Finland under Grant 243/31/2022 "Co-Des: Digital transformation of collaborative powertrain design".

The training and testing computations for this work were performed using the computational resources provided by Aalto University School of Science's “Science-IT” project, through the HPC "Triton".

\nocite{*}
\section{References}
\printbibliography[heading=none]

\appendix
\section{Appendix A: Prompts and Completions}
\label{app:prompts_completions}
This appendix documents the prompt and completion templates used for dataset construction, fine-tuning, and evaluation.

The following prompt was used to produce all training examples containing syntactic errors.

\begin{quote}
The following SysML~v2 code contains compiler-reported syntax errors. 

Repair the code so that it compiles successfully.

Compiler error:

\texttt{\{error\_message\}}

Code:

\texttt{\{bad\_code\}}
\end{quote}

The following prompt was used to generate examples with semantic errors or no errors. It is used only for evaluating baseline models.

\begin{quote}
Check the SysML v2 code below for correctness.

Repair the code if it is incorrect.

If the code is correct, simply report it as correct without rewriting it again.

Code:

\texttt{\{bad\_code\}}
\end{quote}

The following prompt was used to generate examples with semantic errors or no errors. It is employed both for fine-tuning and for evaluating models with additional domain knowledge available.

\begin{quote}
Check the SysML v2 code below for correctness with respect to the given domain constraints.

Repair the code if it is incorrect.

If the code is correct, simply report it as correct without rewriting it again.

Code:

\texttt{\{bad\_code\}}

Domain Rules:

\texttt{\{domain\_rules\_from\_kg\}}
\end{quote}

The following completion was used for examples without errors.

\begin{quote}
CODE STATUS = CORRECT

NO CHANGES REQUIRED
\end{quote}

The following completion was used for examples containing errors.

\begin{quote}
CODE STATUS = INCORRECT

\texttt{\{correct\_code\}}
\end{quote}

The following minimal system prompt was used across all models and variants. It is always added before the main prompt:

\begin{quote}
You are a SysML~v2 expert.
\end{quote}

During fine-tuning, each training instance consists of a prompt–completion pair provided as a single sequence.  
During evaluation and testing, only the prompt is provided to the model.

\end{multicols*}

\end{document}